\documentclass{article}
\usepackage{amsmath,amsthm,amssymb}
\usepackage[pdftex]{graphicx}
\usepackage{url}
\usepackage[hang,small,bf]{caption}
\usepackage[subrefformat=parens]{subcaption}
\usepackage{here}

\theoremstyle{plain}

\renewcommand{\thesection}{\bf{\arabic{section}}}

\begin{document}
\title{Minimising the expectation value of the procurement cost in electricity markets based on the prediction error of energy consumption}
\author{Naoya Yamaguchi, Maiya Hori, Yoshinari Ideguchi}
\date{}

\maketitle

\begin{abstract}
In this paper, 
we formulate a method for minimising the expectation value of the procurement cost of electricity in two popular spot markets: {\it day-ahead} and {\it intra-day}, 
under the assumption that expectation value of unit prices and the distributions of prediction errors for the electricity demand traded in two markets are known. 
The expectation value of the total electricity cost is minimised over two parameters that change the amounts of electricity. 
Two parameters depend only on the expected unit prices of electricity and the distributions of prediction errors for the electricity demand traded in two markets. 
That is, even if we do not know the predictions for the electricity demand, we can determine the values of two parameters that minimise the expectation value of the procurement cost of electricity in two popular spot markets. 
We demonstrate numerically that the estimate of two parameters often results in a small variance of the total electricity cost, and illustrate the usefulness of the proposed procurement method through the analysis of actual data. 
\end{abstract}

\begin{center} Key words; \\ 
minimising the expectation value; electricity markets; prediction error; procurement cost; simulation
\end{center}

\section{\bf{Introduction}}
In recent years, many power exchanges, such as the Japan Electric Power eXchange (JEPX), Amsterdam Power Exchange,  European Power EXchange Spot and PJM Interconnection L.L.C., have been involved in the transfer of energy. 
These power exchanges offer platforms for trading several electricity markets. 
We focus our attention to two popular spot markets: 
{\it day-ahead} and {\it intra-day}. 
The day-ahead market trades electricity one day before delivery, 
whereas the intra-day market trades electricity on the day of delivery.  

In practice, it is important for bidders to make a profits as frequently as possible. 
An accurate forecast of the unit price of electricity may be helpful to increase profits.  Many researchers use statistical models or machine learning to forecast the unit price of electricity 
(e.g., \cite{chinnathambi2016investigation}, \cite{girish2016spot}, \cite{miyauchi2014regression}, \cite{ofuji2007price} and \cite{voronin2013price}.) 
Another method to increase profits is strategic bidding, that is, a simulation of the decision-making process of participants in a market. 
For example, \cite{li2015optimal} studied a methodology to obtain the optimal bidding strategy for generation companies that participate in electric power markets. 
Additionally, \cite{srinivasan2016bidding} considered a coevolutionary approach to estimate individual and cooperative strategies of buyers in a power market.

The total electricity cost also depends on the prediction accuracy of energy consumption because the amount of electricity traded in markets is determined by the result of a prediction. 
As described above, electricity is mainly traded in the day-ahead and intra-day markets.
Trading in these two markets should occur in a particular order: bidders trade the estimated total amount of electricity in the day-ahead market and then trade in the intra-day market if the estimated amount of electricity traded in the day-ahead market is not expected to supply the total demand. 
The unit price of the day-ahead market is usually less than that of the intra-day market. 
Thus, high prediction accuracy in the day-ahead market may increase profits. 

In practice, however, the prediction in the day-ahead market can often be unstable because of the uncertainty of energy consumption. 
As a result, the procurement cost of electricity highly depends on the prediction error of energy consumption.  Therefore, the amount of electricity traded in the two markets must be determined according to the distribution of the prediction error. 
A few researchers have discussed bidding strategies based on the prediction error from a wind power producer's point of view \cite{bathurst2002trading} and \cite{matevosyan2006minimization}. 
These authors minimised the expectation value of the total cost of electricity demand.

However, the methods proposed by \cite{bathurst2002trading} and \cite{matevosyan2006minimization} have three critical problems.  First, the authors estimated the expectation value of the total electricity cost not according to the distribution of the prediction error but according to summary statistics of historical aggregate energy output data, which implies that their methods can lead to an inappropriate expectation value of the total electricity cost.   The second problem is that the variance of the total electricity cost, which is calculated by a distribution of the prediction error, cannot be obtained using the authors' methods.  Third, the authors' methods cannot be applied to multiple electricity markets that are traded in a particular order because their methodologies were developed based on a single market.

To overcome the aforementioned problems, we formulate a method for minimising the expectation value of the procurement cost of electricity. 
Our proposed method can be used when the unit prices (or the expectation values of the unit prices)
of electricity purchased in the two markets and those of supplemental electricity purchased at the time of delivery are provided beforehand (for research on the prediction of the spot price, see the references in \cite{chinnathambi2016investigation} and \cite{miyauchi2014regression}.)  
The distributions of the prediction error of the electricity demand traded in both day-ahead and intra-day markets are taken into account. 
A numerical integration or Monte Carlo simulation can be used to obtain the expectation value of the total electricity cost. 
For the Monte Carlo simulation, the variance of the cost can also be computed. 
The expectation value of the total electricity cost (i.e., the procurement cost) is minimised over two parameters, which modify the amounts of electricity to be purchased in the two markets.  
To show the effectiveness of our proposed procedure, we compute the expectation value and variance of the total electricity cost for various values of tuning parameters when the prediction error follows a normal distribution. 
We demonstrate numerically that our proposed method results in a smaller variance of total electricity cost than a procurement method that does not take the prediction errors into account.  
The usefulness of the proposed procedure is illustrated through the analysis of actual data. 

The remainder of this paper is organised as follows:
in Section~$\ref{markets}$, we explain the day-ahead and intra-day markets and the penalty paid for the purchase of supplemental electricity at the time of delivery.  
In Section~$\ref{method}$, we present our procurement method that prescribes the amounts of electricity purchased from the two markets, 
derive the expected total cost of electricity and express the expectation value of this cost in terms of probability distribution functions of the prediction error.  
In Section~$\ref{normal}$, we report the results of a numerical calculation in which we minimise this expectation value over the two parameters, assuming that the distributions of prediction errors follow normal distributions with mean zero. 
In Section~$\ref{simulation}$, we apply our method to actual data and demonstrate that the procurement cost can be improved by employing our procurement method.

\section{\bf{Electricity markets and the penalty paid for supplemental purchase}}\label{markets}

In this section, we explain the day-ahead market, intra-day market and the penalty for supplemental purchase.

JEPX is one of the most popular electric power exchanges in Japan. 
It operates several electricity exchange markets.
We consider participation in two of these markets: day-ahead and intra-day.
In the day-ahead market, electricity to be delivered on the following day is purchased, and in the
intra-day market, electricity to be delivered on that day is purchased. The delivery schedule of electricity
for a given day is divided into 48 periods that consist of 30 minutes each. The first period is from midnight to 0:30 a.m. and the final period
is from 11:30 p.m. to midnight. For example, if some amount of electricity is procured for the 25th period in
the day-ahead market, then that amount is designated for delivery on the next day between noon to 0:30 p.m, whereas if it is procured
for the 25th period in the intra-day market, then it is designated for delivery on that day between noon and 0:30 p.m.

Suppose we are purchasing electricity for delivery on a given day. We refer to this day as the ``day of delivery."
Then, if for delivery in the $t$-th period,
amount $e_1$ is procured in the day-ahead market on the previous day and amount $e_2$
is procured in the intra-day market on the delivery day, then we have $e_1 + e_2$ of electricity available for delivery
in the $t$-th period on the delivery day. The actual amount of demand for electricity during this period
may be more or less than this amount. We denote this demand by $f(t)$.

Next, we discuss the procedures followed to manage a mismatch between the amount procured and amount demanded
in the cases of $f(t) > e_1 + e_2$ and $f(t) < e_1 + e_2$. When $f(t) > e_1 + e_2$, the surplus amount
of electricity, $e_1 + e_2 - f(t)$, is acquired by an electric company (a general electricity transmission and distribution company).
In the case of $f(t) < e_1 + e_2$, this company provides the (necessary) supplemental amount $f(t) - e_1 + e_2$ and charges a penalty for this service.

In this paper, considering electricity procured for a given period of delivery, we assume the following relations among unit prices:
\begin{align}
\mbox{day-ahead unit price} < \mbox{intra-day unit price} < \mbox{penalty unit price} \; .
\end{align}
However, there are cases in which the penalty unit price is less than one or both market unit prices, and the above inequality does not hold: 
The intra-day unit price is less than the day-ahead unit price.
Assuming that the penalty unit price is the highest price is equivalent to assuming that the market is functioning
properly in accordance with the ``planned-value power balancing system" (see e.g., \cite{Agency}).
Indeed, if the penalty unit price was lower than the market unit prices, then
foregoing participation in the markets and simply procuring all electricity as supplemental purchases and paying the resulting penalties would result in the lowest procurement cost.
However, this would place a large burden on power generation facilities; that is, it would not be
in accordance with the planned-value power balancing system. 
Additionally, relying mainly on the intra-day market for procurement would place a similarly large burden on power generation facilities. 
For these reasons, we assume the relations~(1). 
However, it should be noted that this assumption does not have a significant effect on the results obtained in this work. We only use this
assumption as a condition in the simulations reported below. With this assumption, following the method proposed, most electricity is procured in the day-ahead market.

Assuming that we do not have access to a large storage cell, we are not able to store electricity, and for this
reason, we wish to avoid the scenario in which we procure an amount of electricity that exceeds demand.
However, given the aforementioned relation among unit prices, we also wish to avoid paying penalties. It is necessary
that we attempt to optimally balance these two undesirable scenarios because there is always uncertainty in
our prediction of demand. For this reason, devising a procurement method for minimising the procurement cost
that accounts for this uncertainty is very important for the proper functioning of electricity markets.

\section{\bf{Proposed method}}\label{method}
In this section, we explain the proposed procurement method 
and derive the procurement cost for the total quantity of electricity procured using this method for delivery during the $t$-th period of the delivery day.
Then, we express the expectation value of this cost in terms of probability distributions that represent the errors in the demand predictions.

\subsection{Procurement method}\label{procurement method}
The proposed procurement method takes as its input predictions for demand, $f(t)$, in the $t$-th period on the day of delivery.
These predictions are made at two times: once during the day-ahead market and once during the intra-day market.
With these predictions, using two parameters $A$ and $B$ that account for the effect of the error in these predictions
and the differences among the unit prices, our method yields optimal procurement amounts in the day-ahead and
intra-day markets.

Let $g(t)$ denote the prediction for $f(t)$ made at the time of procurement in the day-ahead market; hereafter,
the ``previous-day prediction."
Then, to hedge against the uncertainty in this prediction, we add some amount $A(t)$ 
and procure the amount $g(t) + A(t)$ in the day-ahead market. The quantity $A(t)$ is to be chosen
in such a manner that optimally accounts for the uncertainty in $g(t)$ and the differences among the unit
prices for the three approaches of procuring electricity.

Next, let $h(t)$ denote the prediction for $f(t)$ made at the time of procurement in the intra-day market; hereafter,
the ``same-day prediction."
Then, in the intra-day market, electricity is procured such that the sum of the amounts procured
in the two markets is $h(t) + B(t)$, where $B(t)$ is a quantity that plays the same role here as $A(t)$ plays in
the day-ahead market. However, in the case that $h(t) + B(t)$ is less than the amount of electricity procured
in the day-ahead market, this is obviously not possible. In this case, no electricity is procured in the intra-day market.

\subsection{\bf{Procurement cost}}\label{derivation}

In this subsection, we derive the total cost of the amount of electricity procured using the method
described in Section~$\ref{method}$.

Let $a$ denote the unit price for the $t$-th period in the day-ahead market.
Because we always consider the $t$-th period in the following discussion, from this point,
we will generally omit the argument $t$ for the various quantities considered. Then, because
the amount procured in the day-ahead market
is $g + A$, the cost of this procurement is
\begin{align}
C_{1} = ( g + A ) a \; .
\end{align}

Next, let $b$ denote the unit price for the $t$-th period in the intra-day market.
Then, because the sum of the amounts procured in the two markets is $h + B$,
if $g + A \le h + B$, then the amount procured in the intra-day market is $h + B - (g + A)$,
whereas if  $g + A \ge h + B$, then the amount procured in the intra-day market is zero.
Thus, the cost of procuring electricity in this market is
\begin{align}
C_{2} = \delta(g + A \leq h + B) (h + B - g - A) b \; ,
\end{align}
where $\delta(*) = 1$ if condition $*$ is satisfied; otherwise, $\delta(*) = 0$ .

$C_{1} + C_{2}$ is the combined cost of electricity procured in the two markets.
To obtain the total procurement cost, we must also calculate the penalty
paid for supplemental electricity, which we denote by $C_3$.
There are two separate scenarios in which we must purchase supplemental electricity:
that in which the condition $g + A \leq h + B \leq f$ is satisfied and that in which the condition
$h + B < g + A \leq f$ is satisfied. Then, with $c$ denoting the unit price for the penalty,
the total amount we pay for the penalty is
\begin{align}
C_{3} &= \delta(g + A \leq h + B \leq f ) (f - h - B) c \nonumber \\ 
&\quad+ \delta(h + B < g + A \leq f) (f - g - A) c \ .
\end{align}

Collecting the above results, the total procurement cost for electricity delivered in the $t$-th period on the
day of delivery is  $C = C_{1} + C_{2} + C_{3}$.

\subsection{Expectation value of the procurement cost}\label{expectation}
In this subsection, we obtain an expression for expectation value $E[C]$ of procurement cost $C$ calculated in Subsection~$\ref{derivation}$
in terms of probability distributions that represent the errors in the demand predictions.

$G = f - g$ and $H = f - h$ are assumed to be random variables, 
and we denote the distribution functions of these variables by $P_{G}(x)$ and $P_{H}(y)$, respectively.
We assume that $f$ and $(a, b, c)$ are independent. 
If the demand $f$ is much smaller than the total trading volume of the electricity market, then this assumption would be valid. 

The expectation value of procuring electricity in the day-ahead market is 
\begin{align}
E[C_{1}] 
&= E[ (g + A) a ] \nonumber \\ 
&= E[a] (E[g] + A) \; .
\end{align}

Next, we obtain the expectation value of procuring electricity in the intra-day market, $C_2$.
For this purpose, we first rewrite $C_2$ in terms of $G$ and $H$:
\begin{align}
C_{2} 
&= \delta(g + A \leq h + B) (h + B - g - A) \nonumber \\ 
&= \delta(A - B \leq G - H) (G - H - A + B) b \; .
\end{align}
From this, we obtain
\begin{align}
E[C_{2}] 
&= E[b] \int_{-\infty}^{\infty} \delta(A - B \leq x) (x - A + B) \int_{-\infty}^{\infty} P_{G}(x + y) P_{H}(y) dy dx \nonumber \\ 
&= E[b] \int_{A-B}^{\infty} (x - A + B) \int_{-\infty}^{\infty} P_{G}(x + y) P_{H}(y) dy dx \nonumber \\ 
&= E[b] \int_{A-B}^{\infty} (x - A + B) P_{G - H}(x) dx \; .
\end{align}

Finally, we obtain the expectation value of the penalty, $C_3$, similarly to that of $C_2$.
First, we rewrite $C_3$ in terms of $G$ and $H$:
\begin{align}
C_{3} 
&= \delta(g + A \leq h + B \leq f) (f - h - B) c \nonumber \\ 
&\quad + \delta(h + B < g + A \leq f) (f - g - A) c \nonumber \\ 
&= \delta(g - f + A - B \leq h - f \leq - B) (H - B) c \nonumber \\ 
&\quad + \delta(h - f - A + B < g - f \leq - A) (G - A)c \nonumber \\ 
&= \delta(B \leq H \leq G - A + B) (H - B) c \nonumber \\ 
&\quad + \delta(A \leq G < H + A - B) (G - A) c \; .
\end{align}
We then immediately obtain
\begin{align}
E[C_{3}] 
&= E[c] \int_{-\infty}^{\infty} \int_{-\infty}^{\infty} \delta(B \leq x \leq y - A + B) (x - B) P_{H}(x) P_{G}(y) dx dy \nonumber \\ 
&\quad + E[c] \int_{-\infty}^{\infty} \int_{-\infty}^{\infty} \delta(A \leq x < y + A - B) (x - A) P_{G}(x) P_{H}(y) dx dy \nonumber \\ 
&= E[c] \int_{A}^{\infty} \int_{B}^{y - A + B} (x - B) P_{H}(x) P_{G}(y) dx dy \nonumber \\ 
&\quad + E[c] \int_{B}^{\infty} \int_{A}^{y + A - B} (x - A) P_{G}(x) P_{H}(y) dx dy \; .
\end{align}

We have thus obtained the following expression for the expectation value of total cost $C$ for
electricity to be delivered in the $t$-th period on the day of delivery:
\begin{align}
E \left[ C \right] 
&= E[C_{1}] + E[C_{2}] + E[C_{3}] \nonumber \\ 
&= E[a] (E[g] + A) \nonumber \\ 
&\quad + E[b] \int_{A-B}^{\infty} (x - A + B) P_{G - H}(x) dx \nonumber \\ 
&\quad \quad + E[c] \int_{A}^{\infty} \int_{B}^{y - A + B} (x - B) P_{H}(x) P_{G}(y) dx dy \nonumber \\ 
&\quad \quad \quad + E[c] \int_{B}^{\infty} \int_{A}^{y + A - B} (x - A) P_{G}(x) P_{H}(y) dx dy \; .
\end{align}

Then, we know that the values of $A$ and $B$ that minimise $E[C]$ depend on $E[a]$, $E[b]$, $E[c]$, $P_{G}(x)$ and $P_{H}(y)$; 
that is, even if we do not know the predictions $g$ and $h$, 
we can determine the values of $A$ and $B$ that minimise $E[C]$ if we know the prediction error distributions $P_{G}(x)$ and $P_{H}(y)$. 

\subsection{$E[C]$ in the case that $G$ and $H$ are normally distributed}\label{normalexp}

In this subsection, we seek the expectation value of the procurement cost $E[C]$ in the case that the
errors in the demand predictions, $G$ and $H$, are normally distributed random variables with mean zero. 

We assume that the difference between the previous-day demand prediction
and actual demand $G$ is normally distributed, with mean zero and variance $\sigma_{1}(t)^{2} = \sigma_{1}^{2}$,
and that 
the difference between the same-day demand prediction
and actual demand $H$ is normally distributed, with mean zero and variance $\sigma_{2}(t)^{2} = \sigma_{2}^{2}$. 
Then, 
$\sigma^{2} = \sigma(t)^{2} = \sigma_{1}(t)^{2} + \sigma_{2}(t)^{2}$ \; , 
the expectation value of  $C_2$ is expressed as 
\begin{align}
E[C_{2}] 
&= E[b] \int_{A-B}^{\infty} (x - A + B) P_{G - H}(x) dx \nonumber \\ 
&= \frac{E[b]}{\sqrt{2 \pi \sigma^{2} }} \int_{0}^{\infty} x \exp \left(- \frac{(x + A - B)^{2}}{2 \sigma^{2}} \right) dx \; ,
\end{align}
and the expectation value of $C_{3}$ is expressed as
\begin{align}
E[C_{3}] 
&= E[c] \int_{A}^{\infty} \int_{B}^{y - A + B} (x - B) P_{H}(x) P_{G}(y) dx dy \nonumber \\ 
&\quad + E[c] \int_{B}^{\infty} \int_{A}^{y + A - B} (x - A) P_{G}(x) P_{H}(y) dx dy \nonumber \\ 
&= \frac{E[c]}{2 \pi \sigma_{1} \sigma_{2}} \int_{A}^{\infty} \int_{B}^{y - A + B} (x - B) \exp{\left( - \frac{x^{2}}{2 \sigma_{2}^{2}} \right)} \exp{\left( - \frac{y^{2}}{2 \sigma_{1}^{2}} \right)} dx dy \nonumber \\ 
&\quad \quad + \frac{E[c]}{2 \pi \sigma_{1} \sigma_{2}} \int_{B}^{\infty} \int_{A}^{y + A - B} (x - A) \exp{\left( - \frac{x^{2}}{2 \sigma_{1}^{2}} \right)} \exp{\left( - \frac{y^{2}}{2 \sigma_{2}^{2}} \right)} dx dy  \; .
\end{align}
We thus have the following result for the expectation value of the procurement cost:
\begin{align}
&E[C] = E[a] (f + A) + \frac{E[b]}{\sqrt{2 \pi \sigma^{2} }} \int_{0}^{\infty} x \exp \left(- \frac{(x + A - B)^{2}}{2 \sigma^{2}} \right) dx \nonumber \\ 
&\quad + \frac{E[c]}{2 \pi \sigma_{1} \sigma_{2}} \int_{A}^{\infty} \int_{B}^{y - A + B} (x - B) \exp{\left( - \frac{x^{2}}{2 \sigma_{2}^{2}} \right)} \exp{\left( - \frac{y^{2}}{2 \sigma_{1}^{2}} \right)} dx dy \nonumber \\ 
&\quad \quad + \frac{E[c]}{2 \pi \sigma_{1} \sigma_{2}} \int_{B}^{\infty} \int_{A}^{y + A - B} (x - A) \exp{\left( - \frac{x^{2}}{2 \sigma_{1}^{2}} \right)} \exp{\left( - \frac{y^{2}}{2 \sigma_{2}^{2}} \right)} dx dy \; .
\end{align}

Because determining an analytical solution to $E[C]$ is difficult, 
we use numerical calculation, such as numerical integration and Monte Carlo simulation.

\section{Numerical calculation of the expectation value of the procurement cost}\label{normal}

In this section, we assume that the errors in the demand predictions are normally distributed random
variables with mean zero. First, in Subsection~$\ref{numerical}$, we report the results of numerical computations in which,
under fixed procurement conditions, we determine the values of parameters $A$ and $B$
that minimise $E[C]$. Next, in Subsection~$\ref{stability}$, we report the results of Monte Carlo simulations
in which we determine the variance of $C$, $V[C]$. We determine that the values of $A$ and $B$ that
minimise $E[C]$ also yield a relatively small value for $V[C]$. Thus, we observe that by carefully choosing
the procurement amounts, we can both minimise the expectation value of the procurement cost and increase its stability.

\subsection{Numerical calculation of $E[C]$}\label{numerical}
With the procurement conditions fixed as
\begin{align}
f = 100, \quad \sigma_{1} = \sqrt{3}, \quad \sigma_{2} = \sqrt{2}, \quad E[a] = 1, \quad E[b] = 2, \quad E[c] = 3 \; ,
\end{align}
we conducted numerical computations in which we determined the expectation value of  $C = C_{1} + C_{2} + C_{3}$. 

First, we explain why we chose the above conditions.

We chose the value $\sqrt{3}$ for the standard deviation $\sigma_{1}$ because, among electric companies,
the target level of precision for the previous-day demand prediction is within $\pm3\%$ \cite[p. 53]{techrepo2016}.
Then, because the precision of the same-day prediction is generally higher, we chose the value $\sigma_{2} = \sqrt{2}$.
Next, for the expectation value of the unit price of the penalty, we chose $E[c] = 3$ because this unit price is often
three times or more greater than the day-ahead unit price \cite[p. 425]{hiroakinagayama2013}.\footnote{In the electricity industry,
the quantity that we refer to as the ``penalty" is often referred to as the ``imbalance fee;" this is the term used in
Ref.~\cite{hiroakinagayama2013}.} Finally, for $E[b]$, we simply used the average of $E[a]$ and $E[c]$. 
In the appendix, we consider variations of these conditions.

\begin{figure}[h]
 \begin{minipage}{0.5\hsize}
  \begin{center}
   \includegraphics[height=4.5cm]{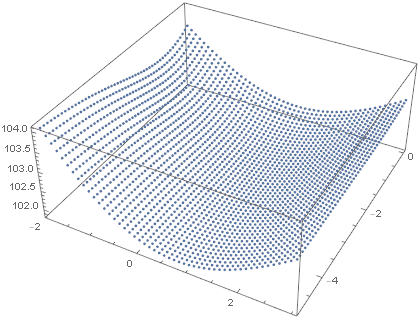}
  \end{center}
  \caption{Expectation value of the procurement cost $E[C]$.}
  \label{plot}
 \end{minipage}
 \begin{minipage}{0.5\hsize}
  \begin{center}
   \includegraphics[height=4.5cm]{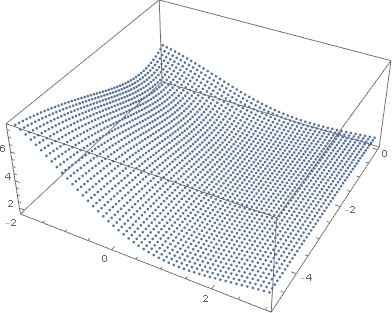}
  \end{center}
  \caption{Variance of the procurement cost $V[C]$.}
  \label{plot2-1}
 \end{minipage}
\end{figure}

Varying $A$ over the range $[-1.9, 3]$ and $B$ over the range $[-4.9, 0]$, we obtained
the plot of $(A, B, E[C])$ shown in Figure~$\ref{plot}$. In this plot, the $A$ axis points
along the 10 o'clock direction and the $B$ axis points along the 1 o'clock direction.
Data points were calculated at intervals of $0.1$ along each axis. For each graph presented in this
paper, the orientation of the $A$ and $B$ axes are the same. Additionally, each graph is plotted using data
obtained for each $(A,B)$ on a grid of mesh size $0.1 \times 0.1$.

Considering the nature of the procurement method, it is clear that sufficiently far
from $(A,B) = (0,0)$, the expectation value of the procurement cost $E[C]$ is
large and increases as we move further away. Thus, we conclude that we can determine the
values of $A$ and $B$ that minimise $E[C]$ by considering only the neighbourhood of $(A,B) = (0,0)$.

Performing the numerical calculation to minimise $E[C]$, in accordance with the conditions stated
above, we found that it was minimised at $(A, B) = (0.6, -2)$, with a minimum value of $E[C] \approx 101.835$.
It is interesting to compare this with the value of $E[C]$ obtained if the amounts procured in the two markets
are simply those given by the predictions, $g$ and $h$ (i.e., the case in which $A = B = 0$.) 
In this case, $E[C] \approx 102.329$.

From the above results, we observe that with the procurement conditions given above, we can reduce $E[C]$ 
by procuring slightly more than the amount of the previous-day demand prediction and slightly less than the
amount of the same-day demand prediction.

\subsection{Stability of $E[C]$}\label{stability}

Generally, even if we have a method to minimise the expectation value of the procurement cost, 
this method is not practically useful when the convergence to the expectation value is slow due to the variance of the cost is too large. 
This is because it entails a large financial risk for companies to use this procurement method when procurement costs continue to be greater than their expected values even if it is a short term. 
In the previous subsection, we reported the values of $A$ and $B$ for which the minimum value
of $E[C]$ is achieved. In this subsection, we investigate the unbiased variance of $C$, $V[C]$, achieved
using these values of $A$ and $B$. Through this investigation, we determine that not only do these
values minimise $E[C]$, but they also result in a relatively small value of $V[C]$.

First, we conducted $10^6$ iterations of a Monte Carlo simulation of the procurement process for each value of $(A,B)$
on a grid of mesh size $0.1 \times 0.1$. From these simulations, we obtained the variance
that corresponded to each such value of $(A,B)$. 
We note that these simulations use equation~(2), (6) and (8). 
Here, we do not use the equation~(13). 

The conditions of the simulations were set as follows:
\begin{align}\label{equ:1}
f = 100, \quad \sigma_{1} = \sqrt{3}, \quad \sigma_{2} = \sqrt{2}, \quad a = 1, \quad b = 2, \quad c = 3 \; .
\end{align}
For the previous-day and same-day demand predictions, we used normal distributions with mean $0$:
\begin{align}
G \sim \mathcal{N}(0, \sigma_{1}^{2}), \quad H \sim \mathcal{N}(0, \sigma_{2}^{2}) \; .
\end{align}

The results of the simulations, $(A, B, V[C])$, are plotted in Fig.~$\ref{plot2-1}$, with values of
$A$ in the range $[-1.9 , 3]$ and values of $B$ in the range $[-4.9, 0]$.

We study the dependence of $C$ on $A$ and $B$, and for this reason,
we express this dependence explicitly by writing $C$ as $C(A,B)$.

In the range of values of $(A,B)$ considered in our simulations, the variance was minimised at $(A,B) = (1, -1.4)$,
and the value was $V[C(1, -1.4)] = 1.693098$.  This value should be compared with the values
of the variance at $(0,0)$ and $(0.6, -2)$, where $E[C]$ is minimised:
$V[C(0, 0)] = 2.879739$; $V[C(0.6, -2)] = 1.821432$. Thus, the variance achieved when electricity was procured by simply following the demand prediction was 1.7 times larger than the minimum value, whereas that achieved for the values of $A$ and $B$ that minimised $E[C]$ was only 1.07 times larger than the minimum value.

\begin{figure}[h]
 \begin{minipage}{0.5\hsize}
  \begin{center}
   \includegraphics[width=6cm]{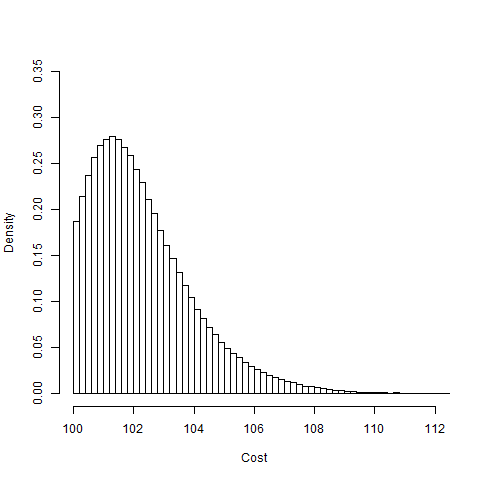}
  \end{center}
  \caption{Histogram for $C(0, 0)$.}
  \label{plot2}
 \end{minipage}
 \begin{minipage}{0.5\hsize}
  \begin{center}
   \includegraphics[width=6cm]{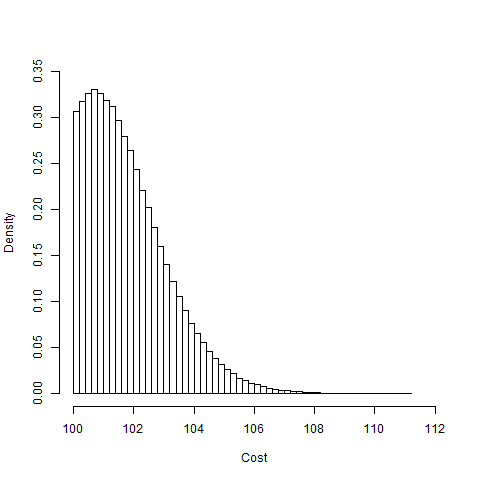}
  \end{center}
  \caption{Histogram for $C(0.6, -2)$.}
  \label{plot1}
 \end{minipage}
\end{figure}

Figures~$\ref{plot2}$ and $\ref{plot1}$ show histograms of the simulation results for $C(A,B)$ obtained in the
cases $(A, B) = (0, 0)$ and $(0.6, -2)$. 

From the above results, we determine that with the proposed method, and under the procurement conditions stated above,
we were able to purchase electricity with a more stable price than in the case that procurement
was made simply in accordance with demand predictions. This effect can be understood
as resulting from a suppression of the penalty paid because of an increase in the amount of electricity procured
in the day-ahead market. Additionally, as can be observed in Fig.~$\ref{plot2-1}$, when $A$ was large, the value of $B$ had almost
no effect on the variance of $C$. This can be understood as resulting from the fact that because there is little difference
between the variances of the previous-day and same-day predictions,
it is usually possible to procure the necessary electricity in the day-ahead market.

\section{\bf{Simulations using actual data}\label{simulation}}
In this section, we report the results of simulations of the proposed procurement method 
using actual data. From these simulations, we determine that this method minimises the procurement cost.  
We use kilowatt-hours (kWh) as the unit of electricity and Japanese yen as the unit of cost. 

The data presented in Table~$\ref{demanddata}$ are the
actual values of demand $f(t,d)$ experienced by a particular facility in Kasuga City, Fukuoka Prefecture,
on 19 weekdays corresponding to the values $d = 1$ to 19, in January 2017 (days 1--3 correspond to January 4 to January 6, days 4--7 correspond to January 10 to January 13, days 8--12 correspond to January 16 to January 20, days 13--17 correspond to January 23 to January 27 and days 18--19 correspond to January 30 to January 31) for the periods $t = 20$ to $t = 26$ (from 9:30 a.m. to 1:00 p.m.) 
We limit the time zone from $t = 20$ to $t = 26$ because the prediction errors of this time zones are large. 

\begin{figure}[h]
  \begin{center}
   \includegraphics[width=12.0cm]{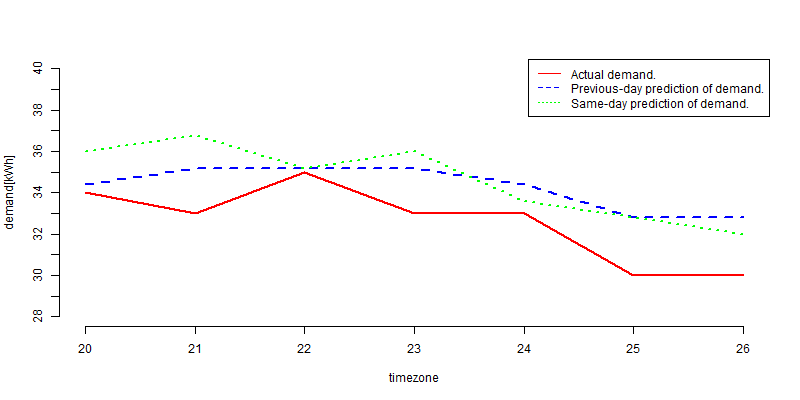}
  \end{center}
  \caption{Actual vs. predicted data ($d = 10$).}
  \label{}
\end{figure}
In actual electricity markets, in the day-ahead market, the unit of traded electricity is 500 kWh \cite[p. 7]{JEPX},
whereas in the intra-day market, it is 50 kWh \cite[p. 12]{JEPX}. 
However, these units of traded electricity are larger than the demand of the facility. 
Therefore, in this paper, we assume the scenario in which the amount of electricity traded in the markets can be freely adjusted and smaller amounts can be traded.

Previous-day predictions $g(t, d)$ and same-day predictions (made 90 minutes before each time period in question)
$h(t, d)$ for the actual demands are presented in Tables~$\ref{previousday}$ and $\ref{sameday}$. \\

In Tables~$\ref{dayahead}$, $\ref{intraday}$ and $\ref{penalty}$, we present 
day-ahead market unit prices $a(t, d)$, intra-day market unit prices $b(t, d)$ and penalty unit prices $c(t,d)$ for the same days and time periods as in the previous tables. The prices listed are the area prices for the Kyushu region.
Additionally, for each intra-day unit price $b(t,d)$, we use the average price during the intra-day market for this time period. 
These data were obtained from the JEPX website on January 31, 2018.

\begin{figure}[h]
  \begin{center}
   \includegraphics[width=12.0cm]{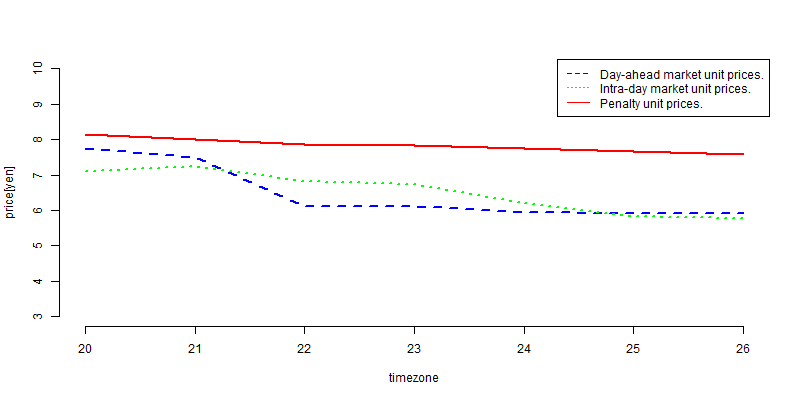}
  \end{center}
  \caption{Unit prices ($d = 1$). }
  \label{}
\end{figure}

To derive the optimal values of $A$ and $B$, we need predictions of the unit prices and error distributions for the demand
predictions. However, we do not presently have a systematic method for generating predictions of unit prices. For example, there is research on price prediction \cite{ofuji2007price}; however, the trading method of JEPX changes frequently, which causes fluctuations in prices, so price prediction does not apply to recent data.

Instead,
we use the following simple prescription for the predictions
of the day-ahead unit price $\hat{a}(t,d)$, intra-day unit price $\hat{b}(t,d)$ and penalty unit price $\hat{c}(t,d)$.
For the time periods 20--24, we use
\begin{align}
\hat{x}(d) = \frac{1}{5} \sum_{20 \leq t \leq 24} x(t, d) \; ,
\end{align}
and for the time periods 25 and 26, we use
\begin{align}
\hat{x}(t) = \frac{1}{19} \sum_{1 \leq d \leq 19} x(t, d) \; ,
\end{align}
where $x$ represents $a$, $b$ and $c$.
We chose these prescriptions for the predictions because, as can be observed from the
Tables~5--7 listing actual prices, the prices during the time periods 20--24 vary significantly from
day to day, whereas the prices in periods 25 and 26 do not. The predictions obtained in this manner
for each time period are presented in Tables~$\ref{dayaheadpred}$,  $\ref{intradaypred}$ and $\ref{penaltypred}$.

For the distribution functions of the errors in the demand predictions, we use normal distributions with mean zero. 
Typically, the demand prediction is performed using normal linear regression modelling. 
In particular, least-squares estimation (LSE) produces a distribution of the prediction error whose mean value is zero 
because the LSE yields an unbiased estimate of the mean of prediction value. 

In practice, the values of $A$ and $B$ are determined before the trade in both day-ahead and intra-day markets. 
Because both $A$ and $B$ depend on the distributions of the prediction error, 
these distributions must be estimated before the trade in the two markets. 
However, the prediction in the intra-day market is performed after the trade in the day-ahead markets; hence, 
the distribution of the prediction error in the intra-day market cannot be obtained before the trade in the day-ahead market. 
Therefore, we need to estimate the distribution of the prediction error without the prediction in the intra-day market to determine the values of $A$ and $B$. 

The estimates of the variances of the prediction error are given as follows: 
For time periods $20$--$24$, to estimate the error variance in the previous-day prediction $\hat{V}_{1}(t, d)$, 
we use the variance of the predicted values of linear regression \cite{fahrmeir2013regression}. 
As an estimator of the variance of the prediction error in the same-day prediction, 
we use 
\begin{align}
\hat{V}_{2}(t) = \frac{1}{19} \sum_{1 \leq d \leq 19} (f(t, d) - h(t, d))^{2} \; .
\end{align}
We present the estimates of the variances obtained in Tables~$\ref{variance for the same}$ and $\ref{variance for the previous}$.

Using the above values for the predictions of unit prices $\hat{a}(t,d)$, $\hat{b}(t,d)$ and $\hat{c}(t,d)$,
and the variances of the errors in the demand predictions $\hat{V}_1(t,d)$ and $\hat{V}_2(t)$, 
we numerically determined the values of $A$ and $B$ that minimised $E[C]$. 
However, in the case $\hat{b} \leq \hat{a}$, we chose $A = 0$,  and in the case $\hat{c} \leq \hat{b}$, we chose $B = 0$ to
maintain the planned-value power balancing system because procuring large amounts of electricity in the
intra-day market or at the time of delivery places a great burden on power generation facilities, and hence is not 
a maintainable market scenario.

Employing the optimised values of $A$ and $B$ that appear in Tables~$\ref{optimize of $A$}$ and $\ref{optimize of $B$}$,
and performing the procurement simulations, we obtained a total procurement cost of
$51,949.95$ yen. We compared this cost with those obtained in two other scenarios.
First, if the previous-day prediction of the
demand was perfect and we procured all the electricity in the day-ahead market, then we would
obtain a total procurement cost of $51,140.72$ yen. Second, using the actual predictions
$g$ and $h$, if we simply procured electricity in accordance with these predictions (i.e., always
using $(A, B) = (0, 0)$), we would have a procurement cost of $52,225.97$ yen. Thus,
the procurement cost achieved with our method was 809.23 yen higher than in the ideal case of perfect demand prediction and 276.02 yen lower than in the case that the ordinary procurement method was followed. 
This result show that the procurement cost can be reduced by about $0.5\%$. 
It is important to reduce the procurement cost by $0.5\%$ for electric power companies. 
 
With the assumption that individual procurements can be made in any amount,
the simulations conducted using actual data demonstrated that our method minimised the procurement cost. 
Because the simulation time was short and the simulation scale was small, the cost reduction was small. 
But, if the simulation time was long and the simulation scale was large, then a significant cost reduction could be expected. 

\section{Conclusion}
We formulated a method for minimising the expected procurement cost of electricity 
and we reported the results of simulations of the proposed procurement method using actual data. 
From these simulations, we found that this method minimised the procurement cost. 
Based on the above discussion, 
to reduce the procurement cost 
it was necessary to estimate not only the accuracy of predictions but also the prediction error distributions. 
In this paper, the prediction error distributions were assumed to be normal distributions, 
but the actual prediction errors did not always become normal distributions. 
To achieve further cost reduction, an estimation of the prediction error distributions is indispensable. 
\clearpage

\thanks{\bf{Acknowledgments}}
We are deeply grateful to Associate Professor Kei Hirose. 
He provided us with helpful comments and suggestions. 
We would also like to thank patent attorney Yuuji Toyomura, 
who provided us with helpful comments and suggestions.

This research was supported by the Japan Science and Technology Agency through its ``Center of Innovation Science and Technology based Radical Innovation and Entrepreneurship Program (COI Program)," and Heisei 29th Energy Research Institute Organization young researcher/doctoral student support program. 

\bibliography{ref.tex}
\clearpage

\appendix
\def\thesection{Appendix:}
\section{Simulations investigating the effect of altering the procurement conditions}\label{conditions}

In this appendix, we consider a variation of the conditions~$(14)$ given in Subsection~$\ref{stability}$. 
Using various combinations of the values of the quantities in the conditions~(14), we again conducted $10^6$ iterations of the Monte Carlo simulation
for a range of values of $(A,B)$. We plot the results for $E[C]$
and $V[C]$, and then discuss how the value of $(A,B)$ that minimises $E[C]$ varies with the variation of
these conditions.

\begin{figure}[h]
 \begin{minipage}{0.5\hsize}
  \begin{center}
   \includegraphics[height=4.5cm]{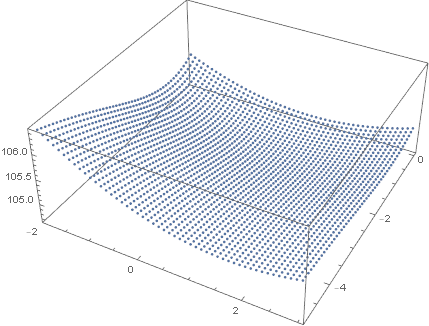}
  \end{center}
  \caption{Expectation value of the procurement cost $E[C]$.}
  \label{plot5-1}
 \end{minipage}
 \begin{minipage}{0.5\hsize}
  \begin{center}
   \includegraphics[height=4.5cm]{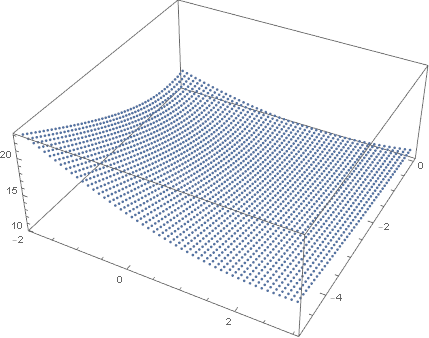}
  \end{center}
  \caption{Variance of the procurement cost $V[C]$.}
  \label{plot5-2}
 \end{minipage}
\end{figure}

First, we consider the case in which the procurement conditions are as in the conditions~(14), except that
$\sigma_{1}$ is changed from $\sqrt{3}$ to $5$. This increase in $\sigma_{1}$ represents
a decrease in the precision of the previous-day demand prediction.
The results of these simulations are displayed in Figs.~$\ref{plot5-1}$ and $\ref{plot5-2}$. 
As in the previous case, data are plotted for $A \in [-1.9, 3]$ and $B \in [-4.9, 0]$.

In the present case, $E[C]$ was minimised at $(A,B) = (0.8, -1)$ and $V[C]$ was minimised
at $(A, B) = (1, -0.4)$. Below, we compare $E[C]$ and $V[C]$ for these two values of $(A, B)$
and $(A,B) = (0,0)$: 
\begin{align*}
&E[C(0.8, -1)] = 104.6559, \quad &&V[C(0.8, -1)] = 10.32363, \\ 
&E[C(1, -0.4)] = 104.7144, \quad &&V[C(1, -0.4)] = 10.15707, \\ 
&E[C(0, 0)] = 104.872, \quad &&V[C(0, 0)] = 10.66363.
\end{align*}

Using the parameter values $A = 0. 6$ and 
$B = -2$, which were the optimal parameter values found using the procurement conditions given in
Section~$\ref{stability}$, as standards for comparison, we found that as the precision of the previous-day demand prediction
decreased, it was advantageous to increase both $A$ and $B$.

\begin{figure}[h]
 \begin{minipage}{0.5\hsize}
  \begin{center}
   \includegraphics[height=4.5cm]{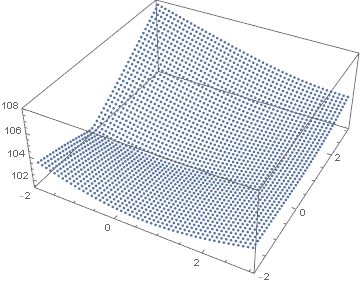}
  \end{center}
  \caption{Expectation value of the procurement cost $E[C]$.}
  \label{plot6-1}
 \end{minipage}
 \begin{minipage}{0.5\hsize}
  \begin{center}
   \includegraphics[height=4.5cm]{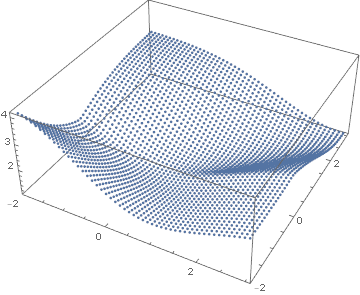}
  \end{center}
  \caption{Variance of the procurement cost $V[C]$.}
  \label{plot6-2}
 \end{minipage}
\end{figure}

Next, we consider the case in which the procurement conditions are as in the conditions~(14), except that $\sigma_{2}$ is
changed from $\sqrt{2}$ to $0.1$. This change represents an increase in the precision of the same-day demand
prediction. The results for $E[C]$ and $V[C]$ obtained
from these simulations are displayed in Figs.~$\ref{plot6-1}$ and $\ref{plot6-2}$. Data are
displayed for $A \in [-1.9, 3]$ and $B \in [-1.9, 3]$.

In this case, $E[C]$ was minimised at $(A, B) = (0.1, -0.1)$, whereas $V[C]$ was minimised
at $(A, B) = (0.1, 0)$. Below, we compare $E[C]$ and $V[C]$ for these values of $(A, B)$
and $(A, B) = (0,0)$:
\begin{align*}
&E[C(0.1, -0.1)] = 101.441, \quad &&V[C(0.1, -0.1)] = 1.101092, \\ 
&E[C(0.1, 0)] = 101.4411, \quad &&V[C(0.1, 0)] = 1.096553, \\ 
&E[C(0, 0)] = 101.4415, \quad &&V[C(0, 0)] = 1.097618. 
\end{align*}

Thus, we observe that when the precision of the same-day demand prediction
improved, it was advantageous to decrease $A$ and increase $B$ from their standard values.

\begin{figure}[h]
 \begin{minipage}{0.5\hsize}
  \begin{center}
   \includegraphics[height=4.5cm]{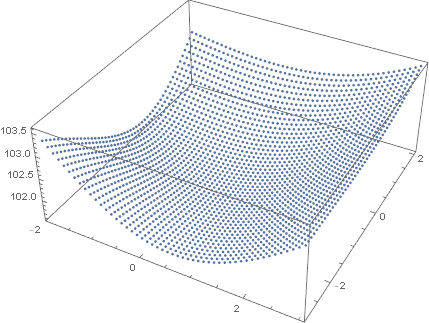}
  \end{center}
  \caption{Expectation value of the procurement cost $E[C]$.}
  \label{plot7-1}
 \end{minipage}
 \begin{minipage}{0.5\hsize}
  \begin{center}
   \includegraphics[height=4.5cm]{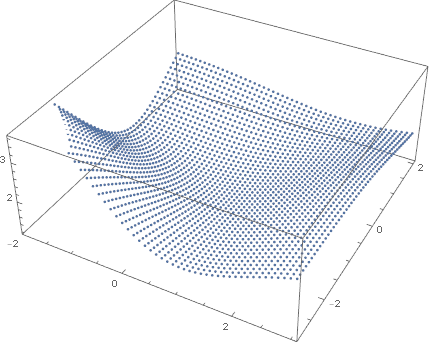}
  \end{center}
  \caption{Variance of the procurement cost $V[C]$.}
  \label{plot7-2}
 \end{minipage}
\end{figure}

As the next scenario, we consider the case in which the procurement conditions are as in the conditions~(14), except that the intra-day unit price $b$ is decreased from 2 to $1.2$.
In Figs.~$\ref{plot7-1}$ and $\ref{plot7-2}$, the results of these simulations for $E[C]$ and $V[C]$
are plotted for $A \in [-1.9 , 3]$ and $B \in [-2.9 , 2]$. 

In this case, $E[C]$ was minimised at $(A, B) = (-0.1, -0.5)$, 
and $V[C]$ was minimised at $(A, B) = (-0.1, 0)$.
Below, we compare $E[C]$ and $V[C]$ for these values of $(A, B)$
and $(A, B) = (0,0)$: 
\begin{align*}
&E[C(-0.1, -0.5)] = 101.5671, \quad &&V[C(-0,1, -0.5)] = 1.24487, \\
&E[C(-0.1, 0)] = 101.608, \quad &&V[C(-0.1, 0)] = 1.178014, \\ 
&E[C(0, 0)] = 101.6139, \quad &&V[C(0, 0)] = 1.179224. 
\end{align*}

Thus, we observe that when the procurement cost in the intra-day market decreased and approached
that in the day-ahead market,
it was advantageous to decrease $A$ and increase $B$ from their standard values.

\begin{figure}[h]
 \begin{minipage}{0.5\hsize}
  \begin{center}
   \includegraphics[height=4.5cm]{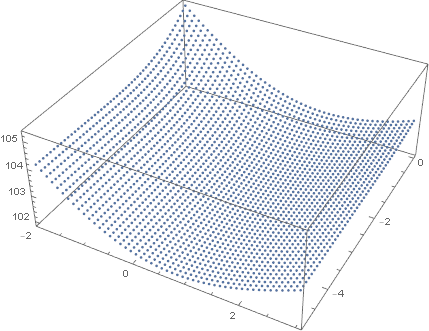}
  \end{center}
  \caption{Expectation value of the procurement cost $E[C]$.}
  \label{plot8-1}
 \end{minipage}
 \begin{minipage}{0.5\hsize}
  \begin{center}
   \includegraphics[height=4.5cm]{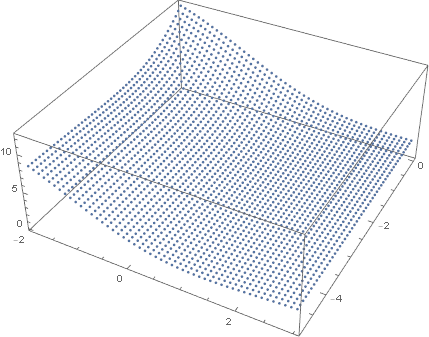}
  \end{center}
  \caption{Variance of the procurement cost $V[C]$.}
  \label{plot8-2}
 \end{minipage}
\end{figure}

As the fourth scenario for comparison, we now consider the case in which the procurement conditions
are as in the conditions~(14), except that intra-day unit price $b$ is increased from 2 to $2.8$.
The results of these simulations are displayed in Figs.~$\ref{plot8-1}$ and $\ref{plot8-2}$ for
$A \in [-1.9 , 3]$ and $B \in [-4.9, 0]$. 

In this case, $E[C]$ was minimised at $(A, B) = (0.7, -3.8)$ and $V[C]$ was
minimised at $(A, B) = (1.2, -2.2)$.  
Below, we compare $E[C]$ and $V[C]$ for these values of $(A, B)$
and $(A, B) = (0,0)$: 
\begin{align*}
&E[C(0.7, -3.8)] = 101.8878, \quad &&V[C(0.7, -3.8)] = 2.1443, \\ 
&E[C(1.2, -2.2)] = 101.9767, \quad &&V[C(1.2, -2.2)] = 1.946507, \\ 
&E[C(0, 0)] = 102.2913, \quad &&V[C(0, 0)] = 2.837869. 
\end{align*}

Thus, we observe that when the intra-day unit price increased and approached the penalty cost,
it was advantageous to increase $A$ and decrease $B$ from their standard values.

\begin{figure}[h]
 \begin{minipage}{0.5\hsize}
  \begin{center}
   \includegraphics[height=4.5cm]{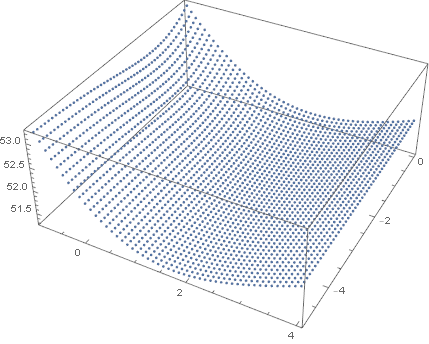}
  \end{center}
  \caption{Expectation value of the procurement cost $E[C]$.}
  \label{plot9-1}
 \end{minipage}
 \begin{minipage}{0.5\hsize}
  \begin{center}
   \includegraphics[height=4.5cm]{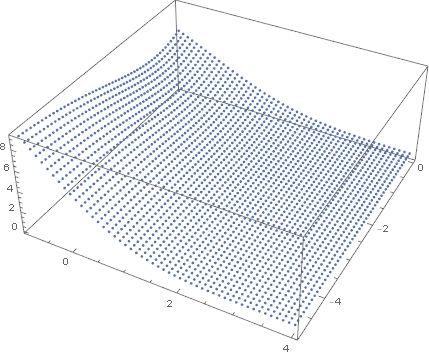}
  \end{center}
  \caption{Variance of the procurement cost $V[C]$.}
  \label{plot9-2}
 \end{minipage}
\end{figure}

As the fifth scenario, we consider the case in which the procurement conditions are as in the conditions~(14), except that the day-ahead unit price $a$ decreases from 1 to 0.5. 
The results of these simulations are displayed in Figs.~$\ref{plot9-1}$ and $\ref{plot9-2}$
for $A \in [-0.9, 3] $ and $B \in [-4.9, 0]$.

In this case, $E[C]$ was minimised at $(A, B) = (1.6, -2.5)$ and $V[C]$ was
minimised at $(A, B) = (3.1, -1.7)$.  
Below, we compare $E[C]$ and $V[C]$ for these values of $(A, B)$
and $(A, B) = (0,0)$: 
\begin{align*}
&E[C(1.6, -2.5)] = 51.2869, \quad &&V[C(1.6, -2.5)] = 1.158393, \\ 
&E[C(3.1, -1.7)] = 51.6331, \quad &&V[C(3.1, -1.7)] = 0.6809917, \\ 
&E[C(0, 0)] = 52.32754, \quad &&V[C(0, 0)] = 4.606727. 
\end{align*}

Thus, we observe that when the day-ahead unit price decreased, it was
advantageous to increase $A$ and decrease $B$ from their standard values.

\begin{figure}[h]
 \begin{minipage}{0.5\hsize}
  \begin{center}
   \includegraphics[height=4.5cm]{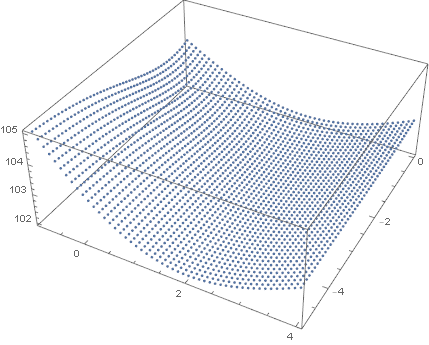}
  \end{center}
  \caption{Expectation value of the procurement cost $E[C]$.}
  \label{plot10-1}
 \end{minipage}
 \begin{minipage}{0.5\hsize}
  \begin{center}
   \includegraphics[height=4.5cm]{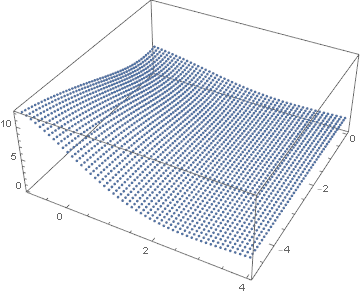}
  \end{center}
  \caption{Variance of the procurement cost $V[C]$.}
  \label{plot10-2}
 \end{minipage}
\end{figure}

As the sixth scenario, we consider the case in which the procurement conditions are as in the conditions~(14), except that penalty unit price $c$ increases from 3 to 3.5.
The results of these simulations are displayed in Figs.~$\ref{plot10-1}$ and $\ref{plot10-2}$
for $A \in [-1.9, 3] $ and $B \in [-4.9, 0]$.

In this case, $E[C]$ was minimised at $(A, B) = (0.8, -1.6)$ and $V[C]$ was
minimised at $(A, B) = (1.2, -1)$.  
Below, we compare $E[C]$ and $V[C]$ for these values of $(A,B)$
and $(A, B) = (0, 0)$:
\begin{align*}
&E[C(0.8, -1.6)] = 101.9741, \quad &&V[C(0.8, -1.6)] = 2.080493, \\ 
&E[C(1.2, -1)] = 102.0595, \quad &&V[C(1.2, -1)] = 1.873814, \\ 
&E[C(0, 0)] = 102.4181, \quad &&V[C(0, 0)] = 3.049496. 
\end{align*}

Thus, we observe that when the penalty unit price increased, it was advantageous to increase
$A$ and decrease $B$ from their standard values.

In Table~$\ref{Changes}$, we summarise the changes undergone by the optimal values of $A$ and $B$ relative to
their standard values because of the various types of changes to the procurement conditions considered above.
\begin{center}
\begin{table}[h]
\begin{tabular}{|c|r|r|}
\hline
& $A$ & $B$ \\ \hline
precision of previous-day demand prediction decreases & increase & increase \\ \hline
precision of same-day demand prediction increases  & decrease & increase \\ \hline
intra-day unit price approaches day-ahead unit price & decrease & increase \\ \hline
intra-day unit price approaches penalty unit price & increase & decrease \\ \hline
day-ahead unit price decreases & increase & decrease \\ \hline
penalty unit price increases & increase & increase \\ \hline
\end{tabular}
\caption{Changes in optimal values of $A$ and $B$ with respect to their standard values.}
\label{Changes}
\end{table}
\end{center}

\begin{center}
\begin{table}[h]
\scalebox{0.75}[0.75]{
\begin{tabular}{l | c c c c c c c c c c c c c c c c c c c}
\hline
$t \backslash d$ & 1 & 2 & 3 & 4 & 5 & 6 & 7 & 8 & 9 & 10 & 11 & 12 & 13 & 14 & 15 & 16 & 17 & 18 & 19 \\ \hline
20 & 28 & 31 & 31 & 33 & 35 & 35 & 35 & 35 & 36 & 34 & 34 & 37 & 36 & 37 & 35 & 35 & 37 & 35 & 33 \\ 
21 & 28 & 30 & 31 & 34 & 34 & 35 & 36 & 35 & 39 & 33 & 33 & 38 & 34 & 36 & 33 & 36 & 37 & 36 & 33 \\ 
22 & 29 & 31 & 33 & 35 & 35 & 35 & 34 & 36 & 38 & 35 & 34 & 38 & 34 & 38 & 34 & 36 & 37 & 35 & 34 \\ 
23 & 30 & 31 & 33 & 34 & 34 & 34 & 34 & 35 & 38 & 33 & 35 & 37 & 34 & 36 & 35 & 35 & 38 & 34 & 32 \\ 
24 & 29 & 30 & 32 & 35 & 32 & 35 & 33 & 35 & 38 & 33 & 33 & 36 & 33 & 36 & 35 & 35 & 36 & 35 & 33 \\ 
25 & 27 & 26 & 28 & 31 & 30 & 32 & 31 & 32 & 34 & 30 & 31 & 33 & 32 & 32 & 33 & 35 & 35 & 31 & 31 \\ 
26 & 28 & 27 & 27 & 33 & 32 & 32 & 33 & 34 & 34 & 30 & 33 & 38 & 32 & 32 & 32 & 34 & 37 & 30 & 32 \\ 
\end{tabular}}
\caption{Data for actual demand.}
\label{demanddata}

\bigskip

\scalebox{0.75}[0.75]{
\begin{tabular}{l | c c c c c c c c c c c c c c c c c c c}
\hline
$t \backslash d$ & 1 & 2 & 3 & 4 & 5 & 6 & 7 & 8 & 9 & 10 & 11 & 12 & 13 & 14 & 15 & 16 & 17 & 18 & 19 \\ \hline
20 & 33 & 32 & 32 & 32 & 32 & 32 & 33 & 33 & 33 & 33 & 34 & 34 & 34 & 34 & 35 & 35 & 36 & 36 & 36 \\ 
21 & 32 & 32 & 31 & 32 & 32 & 32 & 33 & 33 & 33 & 34 & 34 & 34 & 34 & 34 & 35 & 35 & 36 & 36 & 36 \\ 
22 & 31 & 31 & 31 & 32 & 32 & 33 & 33 & 33 & 33 & 34 & 34 & 34 & 35 & 35 & 35 & 36 & 36 & 36 & 36 \\ 
23 & 32 & 31 & 31 & 32 & 32 & 33 & 33 & 33 & 33 & 34 & 34 & 34 & 34 & 34 & 35 & 35 & 36 & 36 & 36 \\ 
24 & 31 & 31 & 31 & 31 & 31 & 32 & 32 & 32 & 33 & 33 & 33 & 33 & 34 & 34 & 34 & 35 & 35 & 36 & 36 \\ 
25 & 29 & 29 & 28 & 28 & 29 & 29 & 30 & 30 & 30 & 31 & 31 & 31 & 31 & 32 & 32 & 33 & 33 & 34 & 34 \\ 
26 & 29 & 30 & 29 & 29 & 30 & 30 & 30 & 31 & 31 & 31 & 32 & 32 & 33 & 33 & 33 & 33 & 34 & 34 & 34 \\ 
\end{tabular}}
\caption{Previous-day predictions of demand.}
\label{previousday}

\bigskip

\scalebox{0.75}[0.75]{
\begin{tabular}{l | c c c c c c c c c c c c c c c c c c c}
\hline
$t \backslash d$ & 1 & 2 & 3 & 4 & 5 & 6 & 7 & 8 & 9 & 10 & 11 & 12 & 13 & 14 & 15 & 16 & 17 & 18 & 19 \\ \hline
20 & 33 & 29 & 31 & 34 & 34 & 34 & 32 & 36 & 34 & 35 & 33 & 32 & 37 & 36 & 35 & 36 & 34 & 37 & 34 \\ 
21 & 32 & 30 & 30 & 34 & 33 & 33 & 33 & 35 & 34 & 36 & 33 & 34 & 36 & 34 & 35 & 34 & 35 & 37 & 36 \\ 
22 & 31 & 30 & 32 & 33 & 34 & 33 & 34 & 36 & 35 & 34 & 35 & 35 & 36 & 35 & 36 & 35 & 36 & 37 & 34 \\ 
23 & 32 & 32 & 31 & 33 & 34 & 33 & 33 & 34 & 35 & 35 & 33 & 36 & 35 & 35 & 34 & 35 & 36 & 36 & 34 \\ 
24 & 31 & 31 & 31 & 33 & 33 & 32 & 34 & 32 & 36 & 32 & 33 & 36 & 33 & 34 & 33 & 36 & 36 & 35 & 34 \\ 
25 & 29 & 29 & 28 & 29 & 30 & 30 & 30 & 31 & 32 & 31 & 30 & 33 & 30 & 34 & 30 & 34 & 35 & 34 & 32 \\ 
26 & 29 & 29 & 29 & 29 & 31 & 31 & 31 & 32 & 34 & 30 & 32 & 33 & 34 & 33 & 32 & 33 & 35 & 34 & 31 \\ 
\end{tabular}}
\caption{Same-day predictions of demand.}
\label{sameday}

\bigskip

\scalebox{0.5}[0.5]{
\begin{tabular}{l | c c c c c c c c c c c c c c c c c c c}
\hline
$t \backslash d$ & 1 & 2 & 3 & 4 & 5 & 6 & 7 & 8 & 9 & 10 & 11 & 12 & 13 & 14 & 15 & 16 & 17 & 18 & 19 \\ \hline
20 & 7.75 & 12.85 & 11.63 & 11.63 & 11.63 & 11.63 & 15.00 & 20.00 & 15.12 & 15.01 & 12.98 & 20.00 & 20.00 & 20.00 & 20.00 & 17.00 & 13.00 & 15.39 & 13.75 \\ 
21 & 7.49 & 12.32 & 11.10 & 11.63 & 10.36 & 11.10 & 11.63 & 17.29 & 14.09 & 13.50 & 11.63 & 16.38 & 17.08 & 15.00 & 17.24 & 11.95 & 12.16 & 20.00 & 11.63 \\ 
22 & 6.11 & 11.93 & 11.10 & 8.90 & 9.70 & 10.33 & 11.63 & 11.63 & 12.20 & 13.00 & 10.56 & 15.00 & 15.00 & 13.00 & 15.00 & 11.75 & 11.63 & 20.00 & 11.62 \\ 
23 & 6.11 & 11.63 & 9.57 & 8.68 & 9.70 & 11.10 & 11.29 & 11.63 & 10.81 & 11.63 & 10.21 & 13.00 & 12.90 & 11.63 & 12.42 & 10.59 & 11.33 & 15.02 & 9.94 \\ 
24 & 5.94 & 8.79 & 9.57 & 8.62 & 9.68 & 9.90 & 10.36 & 11.63 & 11.48 & 11.60 & 10.21 & 13.00 & 12.48 & 11.63 & 12.42 & 10.06 & 10.65 & 13.26 & 8.78 \\ 
25 & 5.91 & 8.68 & 8.16 & 7.40 & 8.27 & 8.44 & 8.27 & 8.83 & 8.97 & 9.07 & 8.68 & 11.63 & 9.89 & 9.89 & 10.06 & 8.05 & 9.89 & 8.93 & 7.42 \\ 
26 & 5.91 & 8.68 & 7.52 & 7.33 & 8.27 & 8.44 & 8.27 & 8.20 & 8.83 & 8.83 & 8.57 & 11.63 & 9.13 & 8.80 & 10.06 & 7.74 & 8.98 & 8.57 & 7.34 \\ 
\end{tabular}}
\caption{Day-ahead market unit prices.}
\label{dayahead}

\bigskip

\scalebox{0.5}[0.5]{
\begin{tabular}{l | c c c c c c c c c c c c c c c c c c c}
\hline
$t \backslash d$ & 1 & 2 & 3 & 4 & 5 & 6 & 7 & 8 & 9 & 10 & 11 & 12 & 13 & 14 & 15 & 16 & 17 & 18 & 19 \\ \hline
20 & 7.11 & 11.58 & 11.33 & 10.27 & 10.97 & 11.05 & 13.42 & 19.34 & 14.27 & 14.57 & 12.41 & 20.92 & 15.68 & 21.93 & 17.14 & 19.19 & 12.52 & 15.17 & 13.64 \\ 
21 & 7.25 & 11.33 & 11.05 & 9.92 & 9.65 & 10.02 & 10.27 & 15.41 & 13.77 & 13.44 & 10.65 & 18.72 & 15.25 & 15.55 & 15.40 & 12.61 & 12.11 & 18.30 & 11.45 \\ 
22 & 6.81 & 11.26 & 10.48 & 8.32 & 9.75 & 9.90 & 10.53 & 11.66 & 12.47 & 12.73 & 9.88 & 17.99 & 13.45 & 14.15 & 13.33 & 11.87 & 11.50 & 18.00 & 10.69 \\ 
23 & 6.74 & 11.46 & 9.74 & 8.32 & 9.61 & 10.04 & 10.49 & 11.24 & 10.43 & 10.44 & 9.93 & 15.50 & 12.74 & 12.03 & 11.73 & 10.03 & 10.79 & 14.66 & 9.43 \\ 
24 & 6.20 & 7.43 & 9.80 & 8.12 & 9.62 & 9.66 & 9.28 & 10.17 & 10.61 & 9.97 & 9.82 & 15.94 & 12.41 & 11.89 & 11.67 & 9.63 & 10.16 & 13.94 & 8.42 \\ 
25 & 5.84 & 8.34 & 9.31 & 7.60 & 8.41 & 8.57 & 8.53 & 9.49 & 9.10 & 8.61 & 8.52 & 13.59 & 9.79 & 9.54 & 10.34 & 7.90 & 9.59 & 7.75 & 8.31 \\ 
26 & 5.76 & 8.20 & 8.51 & 7.62 & 8.34 & 8.41 & 8.55 & 9.23 & 8.68 & 8.45 & 8.47 & 13.64 & 9.42 & 9.46 & 9.88 & 7.89 & 8.86 & 8.62 & 8.04 \\ 
\end{tabular}}
\caption{Intra-day market unit prices.}
\label{intraday}

\bigskip

\scalebox{0.5}[0.5]{
\begin{tabular}{l | c c c c c c c c c c c c c c c c c c c}
\hline
$t \backslash d$ & 1 & 2 & 3 & 4 & 5 & 6 & 7 & 8 & 9 & 10 & 11 & 12 & 13 & 14 & 15 & 16 & 17 & 18 & 19 \\ \hline
20 & 8.16 & 10.29 & 16.48 & 10.79 & 10.17 & 11.75 & 11.20 & 19.87 & 12.36 & 12.95 & 13.22 & 20.35 & 14.36 & 16.13 & 11.66 & 13.08 & 16.50 & 16.74 & 10.87 \\ 
21 & 8.00 & 12.46 & 12.82 & 9.77 & 9.52 & 10.39 & 10.37 & 13.70 & 11.83 & 11.83 & 12.20 & 19.70 & 14.37 & 13.07 & 11.68 & 11.85 & 18.45 & 19.11 & 10.87 \\ 
22 & 7.87 & 10.34 & 10.41 & 9.80 & 9.96 & 10.91 & 10.41 & 12.19 & 11.08 & 9.72 & 11.21 & 20.44 & 13.96 & 11.85 & 10.68 & 10.05 & 13.16 & 13.24 & 9.74 \\ 
23 & 7.83 & 9.82 & 9.74 & 9.07 & 10.28 & 11.85 & 10.91 & 10.92 & 10.91 & 8.97 & 10.95 & 13.55 & 14.05 & 13.10 & 11.58 & 8.89 & 9.47 & 13.52 & 10.21 \\ 
24 & 7.76 & 9.96 & 9.29 & 8.99 & 9.97 & 11.89 & 10.88 & 10.55 & 10.62 & 9.10 & 9.75 & 13.52 & 13.64 & 13.14 & 9.88 & 8.87 & 9.17 & 12.10 & 9.68 \\ 
25 & 7.66 & 10.99 & 8.74 & 8.83 & 9.02 & 10.23 & 10.34 & 9.75 & 9.84 & 9.01 & 8.96 & 10.86 & 11.60 & 12.01 & 9.70 & 8.75 & 9.07 & 8.88 & 9.33 \\ 
26 & 7.59 &11.09 & 8.75 & 8.88 & 9.02 & 9.74 & 11.06 & 9.74 & 9.62 & 8.75 & 8.94 & 10.86 & 11.42 & 11.57 & 9.53 & 8.65 & 8.92 & 8.45 & 8.91 \\ 
\end{tabular}}
\caption{Penalty unit prices.}
\label{penalty}

\end{table}
\end{center}

\begin{center}
\begin{table}[h]

\scalebox{0.5}[0.5]{
\begin{tabular}{l | c c c c c c c c c c c c c c c c c c c}
\hline
$t \backslash d$ & 1 & 2 & 3 & 4 & 5 & 6 & 7 & 8 & 9 & 10 & 11 & 12 & 13 & 14 & 15 & 16 & 17 & 18 & 19 \\ \hline
20--24 & 6.68 & 11.50 & 10.59 & 9.89 & 10.21 & 10.81 & 11.98 & 14.44 & 12.74 & 12.95 & 11.12 & 15.48 & 15.49 & 14.25 & 15.42 & 12.27 & 11.75 & 16.73 & 11.14 \\ 
\end{tabular}
}

\scalebox{0.5}[0.5]{
\begin{tabular}{l | c c c c c c c c c c c c c c c c c c c}
\hline
$t \backslash d$ & 1--19 \\ \hline
25 & 8.76 \\ 
26 & 8.47 \\ 
\end{tabular}}
\caption{Predictions of the day-ahead unit prices.}
\label{dayaheadpred}

\bigskip

\scalebox{0.5}[0.5]{
\begin{tabular}{l | c c c c c c c c c c c c c c c c c c c}
\hline
$t \backslash d$ & 1 & 2 & 3 & 4 & 5 & 6 & 7 & 8 & 9 & 10 & 11 & 12 & 13 & 14 & 15 & 16 & 17 & 18 & 19 \\ \hline
20--24 & 6.82 & 10.61 & 10.48 & 8.99 & 9.92 & 10.13 & 10.80 & 13.56 & 12.31 & 12.23 & 10.54 & 17.81 & 13.91 & 15.11 & 13.85 & 12.67 & 11.42 & 16.01 & 10.73 \\ 
\end{tabular}}

\scalebox{0.5}[0.5]{
\begin{tabular}{l | c }
\hline
$t \backslash d$ & 1--19 \\ \hline
25 & 8.90 \\ 
26 & 8.73 \\ 
\end{tabular}}
\caption{Predictions of the intra-day unit prices.}
\label{intradaypred}

\bigskip

\scalebox{0.5}[0.5]{
\begin{tabular}{l | c c c c c c c c c c c c c c c c c c c}
\hline
$t \backslash d$ & 1 & 2 & 3 & 4 & 5 & 6 & 7 & 8 & 9 & 10 & 11 & 12 & 13 & 14 & 15 & 16 & 17 & 18 & 19 \\ \hline
20--24 & 7.92 & 10.57 & 11.75 & 9.68 & 9.98 & 11.36 & 10.75 & 13.45 & 11.36 & 10.51 & 11.47 & 17.51 & 14.08 & 13.46 & 11.10 & 10.55 & 13.35 & 14.94 & 10.27 \\ 
\end{tabular}}

\scalebox{0.5}[0.5]{
\begin{tabular}{l | c}
\hline
$t \backslash d$ & 1--19 \\ \hline
25 & 9.66 \\ 
26 & 9.55 \\ 
\end{tabular}}
\caption{Predictions of the penalty unit prices.}
\label{penaltypred}
\end{table}
\end{center}

\begin{center}
\begin{table}[h]
\scalebox{0.55}[0.55]{
\begin{tabular}{l | c c c c c c c c c c c c c c c c c c c}
\hline
$t \backslash d$ & 1 & 2 & 3 & 4 & 5 & 6 & 7 & 8 & 9 & 10 & 11 & 12 & 13 & 14 & 15 & 16 & 17 & 18 & 19 \\ \hline
20& 10.48 & 9.41 & 8.46 & 7.64 & 6.95 & 6.38 & 5.94 & 5.63 & 5.44 & 5.37 & 5.44 & 5.63 & 5.94 & 6.38 & 6.95 & 7.64 & 8.46 & 9.41 & 10.48 \\
21 & 13.29 & 11.93 & 10.73 & 9.69 & 8.81 & 8.09 & 7.53 & 7.13 & 6.89 & 6.81 & 6.89 & 7.13 & 7.53 & 8.09 & 8.81 & 9.69 & 10.73 & 11.93 & 13.29 \\
22 & 10.7 & 9.6 & 8.64 & 7.8 & 7.09 & 6.51 & 6.06 & 5.74 & 5.55 & 5.48 & 5.55 & 5.74 & 6.06 & 6.51 & 7.09 & 7.8 & 8.64 & 9.6 & 10.7 \\
23 & 9.29 & 8.34 & 7.5 & 6.78 & 6.16 & 5.66 & 5.27 & 4.99 & 4.82 & 4.76 & 4.82 & 4.99 & 5.27 & 5.66 & 6.16 & 6.78 & 7.5 & 8.34 & 9.29 \\
24 & 10.05 & 9.02 & 8.11 & 7.33 & 6.66 & 6.12 & 5.7 & 5.39 & 5.21 & 5.15 & 5.21 & 5.39 & 5.7 & 6.12 & 6.66 & 7.33 & 8.11 & 9.02 & 10.05 \\
25 & 9.29 & 8.34 & 7.5 & 6.78 & 6.16 & 5.66 & 5.27 & 4.99 & 4.82 & 4.76 & 4.82 & 4.99 & 5.27 & 5.66 & 6.16 & 6.78 & 7.5 & 8.34 & 9.29 \\
26 & 14.69 & 13.19 & 11.86 & 10.72 & 9.74 & 8.95 & 8.33 & 7.89 & 7.62 & 7.53 & 7.62 & 7.89 & 8.33 & 8.95 & 9.74 & 10.72 & 11.86 & 13.19 & 14.69 \\
\end{tabular}}
\caption{Estimated values of the variance of the error distributions for the previous-day demand predictions.}
\label{variance for the same}

\scalebox{0.55}[0.55]{
\begin{tabular}{l | c c}
\hline
$t \backslash d$ & 1--19 \\ \hline
20 & 4.74 \\ 
21 & 5.84 \\ 
22 & 3.05 \\ 
23 & 2.42 \\ 
24 & 2.36 \\ 
25 & 3 \\ 
26 & 4.63 \\ 
\end{tabular}}
\caption{Estimated values of the variance of the error distributions for the same-day demand predictions.}
\label{variance for the previous}
\end{table}
\end{center}

\begin{center}
\begin{table}[h]
\scalebox{0.5}[0.5]{
\begin{tabular}{l | c c c c c c c c c c c c c c c c c c c c}
\hline
$t \backslash d$ & 1 & 2 & 3 & 4 & 5 & 6 & 7 & 8 & 9 & 10 & 11 & 12 & 13 & 14 & 15 & 16 & 17 & 18 & 19 \\  \hline
20 & -5.08 & 0 & 0 & 0 & 0 & 0 & 0 & 0 & 0 & 0 & 0 & -2.12 & 0 & -3.7 & 0 & -5 & 0 & 0 & 0 \\
21 & -5.75 & 0 & 0 & 0 & 0 & 0 & 0 & 0 & 0 & 0 & 0 & -2.42 & 0 & -4.15 & 0 & -5.62 & 0 & 0 & 0 \\
22 & -5.62 & 0 & 0 & 0 & 0 & 0 & 0 & 0 & 0 & 0 & 0 & -2.24 & 0 & -3.73 & 0 & -5.02 & 0 & 0 & 0 \\
23 & -5.31 & 0 & 0 & 0 & 0 & 0 & 0 & 0 & 0 & 0 & 0 & -2.08 & 0 & -3.46 & 0 & -4.68 & 0 & 0 & 0 \\
24 & -5.61 & 0 & 0 & 0 & 0 & 0 & 0 & 0 & 0 & 0 & 0 & -2.17 & 0 & -3.64 & 0 & -4.83 & 0 & 0 & 0 \\
25 & -5.74 & -5.37 & -5.02 & -4.69 & -4.4 & -4.15 & -3.95 & -3.8 & -3.71 & -3.68 & -3.71 & -3.8 & -3.95 & -4.15 & -4.4 & -4.69 & -5.02 & -5.37 & -5.74 \\
26 & -6.33 & -5.91 & -5.55 & -5.21 & -4.9 & -4.65 & -4.44 & -4.27 & -4.18 & -4.15 & -4.18 & -4.27 & -4.44 & -4.65 & -4.9 & -5.21 & -5.55 & -5.91 & -6.33 \\
\end{tabular}}
\caption{Optimal values of $A$.}
\label{optimize of $A$}

\bigskip

\scalebox{0.5}[0.5]{
\begin{tabular}{l | c c c c c c c c c c c c c c c c c c c c}
\hline
$t \backslash d$ & 1 & 2 & 3 & 4 & 5 & 6 & 7 & 8 & 9 & 10 & 11 & 12 & 13 & 14 & 15 & 16 & 17 & 18 & 19 \\  \hline
20 & -2.71 & 0 & -4.53 & -5.21 & -7.3 & -4.88 & 0 & 0 & 0 & 0 & -5.48 & 0 & -6.28 & 0 & 0 & 0 & -4.02 & 0 & 0 \\
21 & -2.99 & 0 & -4.99 & -5.74 & -8.96 & -5.32 & 0 & 0 & 0 & 0 & -6.03 & 0 & -7.86 & 0 & 0 & 0 & -4.48 & 0 & 0 \\
22 & -2.04 & 0 & -3.22 & -3.79 & -5.57 & -3.46 & 0 & 0 & 0 & 0 & -4 & 0 & -5.94 & 0 & 0 & 0 & -2.93 & 0 & 0 \\
23 & -1.81 & 0 & -2.84 & -3.32 & -4.86 & -3.02 & 0 & 0 & 0 & 0 & -3.45 & 0 & -4.84 & 0 & 0 & 0 & -2.54 & 0 & 0 \\
24 & -1.77 & 0 & -2.77 & -3.19 & -4.77 & -2.89 & 0 & 0 & 0 & 0 & -3.34 & 0 & -4.52 & 0 & 0 & 0 & -2.46 & 0 & 0 \\
25 & -2.6 & -2.63 & -2.66 & -2.69 & -2.73 & -2.76 & -2.79 & -2.81 & -2.83 & -2.84 & -2.83 & -2.81 & -2.79 & -2.76 & -2.73 & -2.69 & -2.66 & -2.63 & -2.6 \\
26 & -3.2 & -3.23 & -3.27 & -3.32 & -3.36 & -3.4 & -3.44 & -3.47 & -3.49 & -3.5 & -3.49 & -3.47 & -3.44 & -3.4 & -3.36 & -3.32 & -3.27 & -3.23 & -3.2 \\
\end{tabular}}
\caption{Optimal values of $B$.}
\label{optimize of $B$}
\end{table}
\end{center}

\end{document}